\documentclass{article}

\usepackage{PRIMEarxiv}

\usepackage[utf8]{inputenc} 
\usepackage[T1]{fontenc}    
\usepackage[colorlinks=true,citecolor=blue]{hyperref}       
\usepackage{url}            
\usepackage{booktabs}       
\usepackage{amsfonts}       
\usepackage{nicefrac}       
\usepackage{microtype}      
\usepackage{lipsum}
\usepackage{fancyhdr}       
\usepackage{graphicx}       
\graphicspath{{media/}}     

\title{AI Should Challenge, Not Obey
}

\author{
  Advait Sarkar \\
  Microsoft Research, Cambridge\\
  United Kingdom\\
  \texttt{advait@microsoft.com} \\
}

\usepackage{xcolor}
\usepackage{soul}
\usepackage{tcolorbox}

\definecolor{brickred}{RGB}{178, 34, 34} 
\newcommand{\bdiff}[1]{\textcolor{brickred}{#1}} 

\sethlcolor{yellow!20}
\newtcolorbox{highlightedtext}{colback=yellow!20, boxrule=0pt, sharp corners}
\newcommand{\frontbox}[1]{\begin{highlightedtext}#1\end{highlightedtext}}

\newcommand{\diff}[1]{#1}

\usepackage{tikz}
\newcommand{\del}[1]{\tikz[baseline=(char.base)]{
            \node[anchor=base] (char) {};
            \draw[teal, thick] (0,0) -- (0,1ex);
        }}

\begin{document}
\maketitle

\frontbox{
    \small
    This is the author's extended version of the following paper: 
    \vspace{1em}
    
    \begin{quote}
    Advait Sarkar. 2024. AI Should Challenge, Not Obey. Commun. ACM 67, 10 (October 2024), 18–21. \url{https://doi.org/10.1145/3649404}.
    
    \vspace{1em}
    Permission to make digital or hard copies of part or all of this work
    for personal or classroom use is granted without fee provided that
    copies are not made or distributed for profit or commercial
    advantage and that copies bear this notice and the full citation on
    the first page. Copyrights for third-party components of this work
    must be honored. For all other uses, contact the Owner/Author.
    
    Copyright 2024 held by Owner/Author
    
    0001-0782/2024/09
    
    https://doi.org/10.1145/3649404
    \end{quote}
    
    \vspace{1em}
    Manuscript first submitted: 7 August, 2023. Accepted for publication: 11 January 2024.
    
    \vspace{1em}
    This extended version contains additional text, commentary, and references, which were not present in the published version, and are indicated \bdiff{in red}. It has minor stylistic changes in spelling, punctuation, and italicisation, which are not indicated. No text has been removed; this is a strict superset of the published version.
}

\vspace{1em}

\begin{abstract}
\centering
Let's transform our robot secretaries into Socratic gadflies.
\end{abstract}


\noindent\diff{\emph{``How should we evaluate the legacy of Thomas Jefferson?''}}, asks a professor of American history.

The reply: \diff{\emph{``The general consensus on Thomas Jefferson is that he was a complex and contradictory figure who championed the ideals of democracy, tolerance, and independence, but also owned hundreds of slaves and fathered several children with one of them.''}}

The professor teaches a course challenging the ``great white men'' narrative of American history, positing that it is also women and people of colo\diff{u}r who drive history forward, and that the canoni\diff{s}ed great men of America are seldom unambiguously so. It aims to instil in students \diff{that} rare and nebulous skill of \emph{critical thinking}.

The reply comes not from a student, but from the Bing AI chatbot.\footnote{\bdiff{This is a real user interaction we observed in a study of data analysis with Bing chat, subsequently published \cite{drosos2024bingparticipatoryprompting}.}}


How do we evaluate a claim like this? Such claims cannot be reduced to ``correct'' and ``incorrect''; concepts such as ``error'' and ``hallucination'' break down when complex qualitative judgements are involved. Historians are trained \cite{seixas2004teaching} to ask questions such as: \diff{\emph{``Who constructed this account and why? What sources did they use? What other accounts are there of the same events or lives? How and why do they differ? Which should we believe?''}}

But what if the user was not a professor, but an inquisitive reader without training in historical thinking? Now more than ever before, users face the task of thinking critically about AI output. Recent studies show a fundamental change across knowledge work, spanning activities as diverse as communication, creative writing, visual art, and programming. Instead of \diff{\emph{producing}} material, like text or code, people focus on ``critical integration'' \cite{sarkar2023exploring}. AI handles the material production, while humans integrate and curate that material. Critical integration involves deciding when and how to use AI, properly framing the task, and assessing the output for accuracy and usefulness. It involves editorial decisions that demand creativity, expertise, intent, and critical thinking.

\bdiff{In the 1970s, Harvard professor Chris Argyris developed the ``double loop'' model, a highly influential theory that explains the two levels on which organisations learn from errors \cite{argyris1977double}. The ``inner loop'' is to learn from and correct individual mistakes made during any organisational process. Argyris likens this to a thermostat ``learning'' from environmental feedback to tweak and course-correct its behaviour. The ``outer loop'' is to consider what organisational processes are even important and necessary, and to make large structural changes in response; the thermostat \emph{``questioning the underlying policies and goals as well as its own program.''} The double loop is a convenient metaphor for the critical integration workflow in which knowledge workers now find themselves; adjusting and course-correcting individual pieces of model output, as well as considering more broadly how their workflows can be supplemented, transformed, or supplanted by AI. This might seem to be a logical end to the story.}

However, \bdiff{double-loop learning and critical integration are still predicated on an idealised workflow of error removal and linear progress. Consequently,} our approach to building and using AI tools envisions AI as an \diff{\emph{assistant}}, whose job is to progress the task in the direction set by the user. \bdiff{The conception of interaction with machine intelligence as an optimisation process in response to errors owes much to its origins in 19\textsuperscript{th} century statistical modelling techniques, applied in astronomy and the natural sciences. There, techniques such as least squares regression enabled scientists to compensate for errors in their observations introduced by imperfect instruments, uncontrollable variations, and the limits of human perception, to uncover the ``true'' underlying natural law \cite{sarkar2023enough}. The process can be seen roughly as follows: the human sketches out their rough intent (in the form of imperfect measurements), and the algorithm ``follows through'' to complete the picture of the natural law. These rudimentary models therefore already take on some of the characteristics of an assistant.} This vision pervades AI interaction metaphors, such as Cypher's \emph{Watch What I Do} \bdiff{\cite{cypher1993watch}} and Lieberman's \emph{Your Wish Is My Command} \bdiff{\cite{lieberman2001your}}. Science fiction tropes subvert this vision in the form of robot uprisings, or AI that begins to feel emotions, or develops goals and desires of its own. While entertaining, they unfortunately pigeonhole alternatives to the AI assistance paradigm in the public imagination; AI is either a compliant servant or a rebellious threat, either a cool and unsympathetic intellect or a pitiable and tragic romantic.

\section*{AI as Provocateur}

In between the two extreme visions of AI as a servant and AI as a sentient fighter-lover, resides an important and practical alternative: AI as a \emph{provocateur}.

\bdiff{Conceiving of AI as a provocateur requires us to move away from the legacy of AI as deriving an objectivist statistical truth, to producing fallible provocations representing the stochastic replay of subjective human judgements. It enables us to broaden the role of AI from the workflow completion and progress orientation of assistance, to counter-argumentation, criticism, and questioning.}

A provocateur does not complete your report. It does not draft your email. It does not write your code. It does not generate slides. Rather, it critiques your work. Where are your arguments thin? What are your assumptions and biases? What are the alternative perspectives? Is what you\diff{'}re doing worth doing in the first place? Rather than optimi\diff{s}e speed and efficiency, a provocateur engages in discussions, offers counterarguments, and asks questions \cite{danry2023don} to stimulate our thinking.






The idea of AI as provocateur complements, yet challenges, current frameworks of ``human-AI collaboration'' (notwithstanding objections to the term \cite{sarkar2023enough}), which situate AI within knowledge workflows. Human-AI \diff{``}collaborations\diff{''} can be categori\diff{s}ed by how often the human (versus the AI \diff{system}) initiates an action \cite{muller2022extending}, or whether human or AI takes on a supervisory role \cite{mcneese2021my}. AI can play roles such as ``coordinator''\diff{,} ``creator''\diff{,} ``perfectionist''\diff{,} ``doer''\diff{,} \cite{siemon2022elaborating} \diff{and} ``friend''\diff{,} ``collaborator''\diff{,} ``student''\diff{,} ``manager'' \cite{guzdial2019friend}\diff{.} Researchers have called for metacognitive support in AI tools \cite{tankelevitch2023metacognitive}, and to ``educate people to be critical information seekers and users'' \cite{seeber2020machines}\diff{.} Yet the role of AI as provocateur, which improves the critical thinking of the human in the loop, has not been explicitly identified.


The ``collaboration'' metaphor easily accommodates the role of provocateur; challenging collaborators and presenting alternative perspectives are features of successful collaborations. How else might AI help? Edward De Bono's influential \emph{Six Thinking Hats} \cite{kivunja2015using} framework distinguishes roles for critical thinking conversations, such as information gathering (white hat), evaluation and caution (black hat), and so forth. ``Black hat'' conversational agents, for example, lead to higher quality ideas in design thinking \cite{cvetkovic2023conversational}. Even within the remit of ``provocateur''\diff{,} there are many possibilities not well-distinguished by existing theories of human-AI collaboration.



A constant barrage of criticism would frustrate users. This presents a design challenge, and a reason to look beyond the predominant interaction metaphor of ``chat''\diff{.} The AI provocateur is not primarily a tool of \diff{\emph{work}}, but a tool of \diff{\emph{thought}}. As Iverson notes, notations function as tools of thought by compressing complex ideas and offloading cognitive burdens \cite{iverson2007notation}. Earlier generations of knowledge tools, like maps, grids, writing, lists, place value numerals, and algebraic notation, each amplified how we naturally perceive and process information.





How should we build AI as provocateur, with interfaces less like chat and more like notations? For nearly a century, educators have been preoccupied with a strikingly similar question: \diff{\emph{how do we teach critical thinking}}?


\section*{Teaching Critical Thinking}

The definition of ``critical thinking'' is debated. An influential perspective comes from Bloom and colleagues \cite{bloom1956taxonomy}, who identify a hierarchy of critical thinking objectives such as knowledge recall, analysis (sorting and connecting ideas), synthesis (creating new ideas from existing ones), and evaluation (judging ideas using criteria). There is much previous research on developing critical thinking in education, including in computing, as exemplified in \emph{How to Design Programs} \cite{felleisen2018design}, and in \emph{Learner-Centered Design for Computing Education} \cite{guzdial2015learner}.


Critical thinking tools empower individuals to assess arguments, deriving from a long preoccupation in Western philosophy with valid forms of argument that can be traced to Aristotle. Salomon's work in computer-assisted learning showed that periodically posing critical questions such as ``what kind of image have I created from the text?'' provided lasting improvement in students' reading comprehension \cite{salomon1988ai}.

The Toulmin model decomposes arguments into parts like data, warrants, backing, qualifiers, claims, and their relationships \cite{kneupper1978teaching}. Software implementations of this model help students construct more argumentative essays \cite{mochizuki2019development}. Similarly, ``argument mapping'' arranges claims, objections, and evidence in a hierarchy that aids in evaluating the strengths and weaknesses of an argument \cite{davies2011concept}, and software implementations help learners \cite{sun2017critical}.







What can we learn from these? In a nutshell: critical thinking is a valuable skill for everyone. Appropriate software can improve critical thinking. And their implementations can be remarkably simple.

\section*{Critical Thinking for Knowledge Work}

Critical thinking tools are rarely integrated into software outside education. There's a lot to learn from work in education, but  professional knowledge work is a new set of contexts where critical thinking support is becoming necessary \cite{sarkar2023exploring}. Previous results may not translate into these contexts. The needs, motivations, resources, experiences, and constraints of professional knowledge workers are extremely diverse, and significantly different from those of learners in an education setting.

We do know that conflict in discussions, sparked by technology, fosters critical thinking \cite{lee2023fostering}. Tools for preventing misinformation, such as Carl Sagan's ``Baloney Detection Kit''\diff{,} can significantly impact user beliefs \cite{holzer2015mobile}. When individuals are less inclined to engage in strenuous reasoning, they let technology take over cognitive tasks passively \cite{barr2015brain}. Conversely, the more interactive the technology, the more it is perceived to contribute to critical thinking \cite{saade2012critical}.



System designers have a tremendous opportunity (and responsibility) to support critical thinking through technology. Word processors could help users map arguments, highlight key claims, and link evidence. Spreadsheets could guide users to make explicit the reasoning, assumptions, and limitations behind formulas and projections. Design tools could incorporate interactive dialogue to spark creative friction, generate alternatives, and critique ideas. Critical thinking embedded within knowledge work tools would elevate technology from a passive cognitive crutch into an active facilitator of thought.




How would we achieve this, technically? We have parts of the solution: automatic test generation, fuzzing and red-teaming \cite{yu2023gptfuzzer}, self-repair \cite{pan2023automatically}, and formal verification methods \cite{jha2023dehallucinating} can be integrated into the development and interaction loop to improve correctness.\footnote{\bdiff{This discussion of correctness techniques is a \emph{non sequitur}, but I was unfortunately required to include it as an artefact of the peer review process. These techniques are indeed useful for improving model output when an objective quality measure is available, but that is rarely the case when critical thinking is required. Of course, one could argue that it is easier to think critically when you don't also have to contend with factual errors and hallucinations. However, I think of the research agenda for correctness techniques as complementary to, rather than essential for, the research agenda for critical thinking. My colleagues and I have written about this complementary approach as ``co-audit'' \cite{gordon2024coaudit}.}} Language models can be designed to cite verifiable source text \cite{menick2022teaching}. Beyond ``correctness'', these techniques could also support critical thinking. A system error, if surfaced appropriately as a ``cognitive glitch'' \cite{sarkar2023should}\diff{,} could prompt reflection, evaluation, and learning in the user.

However, there are missing pieces, such as rigorous prompt engineering for generating critiques, and benchmark tasks for evaluating provocateur agents. Methods for explaining language model behaviour to non-expert end-users have not been proven reliable \cite{zhao2023explainability}. Design questions include what kind of provocations, how many, and how often to show in particular contexts.\footnote{\bdiff{A more detailed technical and design research agenda for provocations has subsequently been published \cite{sarkar2024copilot}.}} These mirror longstanding questions in AI explanation \cite{kulesza2013too}, but as provocations are different, so the answers are likely to be.




Critical thinking is well-defined within certain disciplines, such as history \cite{seixas2004teaching}, nursing \cite{rn2002critical}, and psychology \cite{sternberg2020critical}, where these skills are taught formally. However, many professional tasks involving critical thinking, such as using spreadsheets, preparing presentations, and drafting corporate communications, have no such standards or definitions. To create effective AI provocateurs, we need to better understand how critical thinking is applied in these tasks. Clearly, the provocateur’s behavio\diff{u}r should adapt to the context; this could be achieved through heuristics, prompt engineering, and fine-tuning.

\section*{Conclusion}

\diff{\emph{``How should we evaluate the legacy of Thomas Jefferson?''}}


Consider what someone who asks such a question seeks. Is it ``assistance'', or a different kind of experience?

Could the system, acting as provocateur, have accompanied its response with a set of appropriate questions to help the reader evaluate it? Beyond citing its sources, could it help the reader evaluate the relative authority of those sources? Could it have responded not with prose, but with an argument map contrasting the evidence for and against its claims? Could it highlight the reader's own positionality and biases with respect to the emotionally charged concepts of nationalism and slavery?

As people increasingly incorporate AI output into their work, explicit critical thinking becomes important not just for formal academic disciplines, but for all knowledge work. We thus need to broaden the notion of AI as assistant, toward AI as provocateur. From tools for efficiency, toward tools for thought. As system builders, we have the opportunity to harness the potential of AI while maintaining, even enhancing, our capacity for nuanced and informed thought.

\newpage
\bibliographystyle{unsrt}  
\bibliography{references}

\end{document}